\documentclass[aps,prb,reprint,superscriptaddress]{revtex4-1}
\usepackage{graphicx}

\begin{document}

%\preprint{}

\title{New Group V Elemental Bilayers: A Tunable Structure Model with 4,6,8-atom Rings}


\author{Xiangru Kong}
\email[]{kongxru@pku.edu.cn}
%\thanks{}
\affiliation{International Center for Quantum Materials, School of Physics, Peking University, Beijing, 100871, China and
Collaborative Innovation Center of Quantum Matter, Beijing 100871, China}

\author{Linyang Li}
\email[]{linyang.li@uantwerpen.be}
%\thanks{}
\affiliation{Department of Physics, University of Antwerp, Groenenborgerlaan 171, B-2020 Antwerp, Belgium}

\author{Ortwin Leenaerts}
%\thanks{}
\affiliation{Department of Physics, University of Antwerp, Groenenborgerlaan 171, B-2020 Antwerp, Belgium}

\author{Xiong-jun Liu}

%\thanks{}
\affiliation{International Center for Quantum Materials, School of Physics, Peking University, Beijing, 100871, China and
Collaborative Innovation Center of Quantum Matter, Beijing 100871, China}

\author{Fran\c{c}ois M. Peeters}
%\thanks{}
\affiliation{Department of Physics, University of Antwerp, Groenenborgerlaan 171, B-2020 Antwerp, Belgium}


\date{\today}

\begin{abstract}
Two-dimensional (2D) group V elemental materials have attracted widespread attention due to their nonzero band gap while displaying high electron mobility. Using first-principles calculations, we propose a series of new elemental bilayers with group V elements (Bi, Sb, As). Our study reveals the dynamical stability of 4, 6, and 8-atom ring structures, demonstrating their possible coexistence in such bilayer systems. The proposed structures for Sb and As are large-gap semiconductors that are potentially interesting for applications in future nanodevices. The Bi structures have nontrivial topological properties with a large and direct nontrivial band gap. The nontrivial gap is shown to arise from a band inversion at the Brillouin zone center due to the strong intrinsic spin-orbit coupling (SOC) in Bi atoms. Moreover, we demonstrate the possibility to tune the properties of these materials by enhancing the ratio of 6-atom rings to 4 and 8-atom rings, which results in wider nontrivial band gaps and lower formation energies. 

\end{abstract}

%\keywords{}

\maketitle

\section{Introduction}
The story of 2D materials begins with the successful exfoliation of graphene from graphite.\cite{Novoselov2004} 2D materials are usually defined as crystalline materials consisting of a single or few layers of atoms. The unusual physical properties, caused by dimensional restrictions, lead researchers to study these materials for possible use in applications and future nanodevices.\cite{Kim2015} The search for other 2D materials besides graphene is an on-going field of research. In analogy to graphene, other group IV elements also form 2D hexagonal structures, such as silicene,\cite{Feng2012a,PhysRevLett.108.155501,Chiappe2012,Chiappe2014} germanene,\cite{Bianco2013,Derivaz2015a} and stanene,\cite{Zhu2015a} and have been successfully synthesized on different substrates. Similar structures could also be observed for the 2D group V elemental structures. In theory, buckled hexagonal honeycomb bilayers of group V elements are also stable and favorable in energy.\cite{zhang2015atomically} For example, hexagonal Bi(111) bilayers\cite{PhysRevLett.114.066402,Drozdov2014a,PhysRevB.90.165412,PhysRevLett.107.136805} have been experimentally synthesized on Bi$_2$Te$_3$ and Bi$_2$Se$_3$ surfaces.\cite{PhysRevLett.109.016801,PhysRevLett.107.166801,PhysRevLett.109.227401} In this connection, the successful growth of single layer blue phosphorus has attracted widespread attention to the group V elemental bilayers due to their nonzero band gap and high electron mobility.\cite{doi:10.1021/acs.nanolett.6b01459} 
There have also been many suggestions for other 2D stable carbon allotropes,\cite{Zhang2015,PhysRevLett.108.225505} such as phagraphene \cite{doi:10.1021/acs.nanolett.5b02512} and graphyne,\cite{PhysRevB.86.115435,Wang201465} and some of them have been successfully created or can be found as defects in graphene. The physical properties of such crystalline materials mainly originate from the underlying symmetry of the crystal structure. Therefore, it is interesting to study 2D crystal structures with different symmetries. Recently, some theoretical works have studied 2D group V structures with 4-atom and 8-atom rings on a square lattice. \cite{2053-1583-2-4-045010,Li2016}  However, the formation energy of these bilayer structures is relatively high which makes it difficult to realize them in experiments. It is thus an interesting question how such materials can be made more stable.%This leaves us a question if there exists other, more stable, 2D allotropes with group V elements? 

%However, the discovery of 2D materials consist of the  the group-V elements is also amazing. Firstly, the Bi (111) bilayer \cite{PhysRevLett.114.066402,Drozdov2014a,PhysRevB.90.165412,Liu2011} which have the buckled honeycomb structures grown on Bi$_2$Te$_3$ or Bi$_2$Se$_3$ \cite{PhysRevLett.109.016801,PhysRevLett.107.166801,PhysRevLett.109.227401} have been synthesized in experiments. Importantly, the discovery of black phosphorus (BP) \cite{Li2014b} has attracted more attention focused on the group-V elements because of the nonzero band gap while displaying high electron mobility. Up to now, except for BP, other 2D materials with the group-V elements such as the allotrope of BP blue phosphorus \cite{Zhang2016b} have also been reported. The fact is that carbon has many other 2D allotropes \cite{Zhang2015,Liu2012} except for graphene while it is rare for other elements. However, the physical properties of crystalline materials origin from the underlying symmetry in the spatial structures, and the effort in searching for dynamical stable 2D materials with group-V elements to discovery new physical phenomena is deserved. Recently, some theoretical articles have studied the dynamical stable 2D group-V structures with 4-atom and 8-atom rings. \cite{2053-1583-2-4-045010,Li2016} However, the energy in the 2D group-V structures with 4-atom and 8-atom rings could be very high and make it difficult in the realization in experiments. This leaves us a question if there exists other more dynamical stable 2D allotropes with group-V elements? 

One of the most intriguing properties of some 2D materials is their nontrivial band topology. 2D topological insulators with time-reversal (TR) symmetry, also known as quantum spin Hall (QSH) insulators, are a very important set of 2D materials.\cite{RevModPhys.83.1057,RevModPhys.82.3045,PhysRevLett.96.106802,PhysRevLett.95.226801} Graphene was the first proposal for such a topological insulator, but its negligible nontrivial band gap makes it impossible to observe the QSH effect.\cite{PhysRevLett.95.226801,PhysRevB.75.041401} In experiment, the QSH effect has been obeserved in HgTe/CdTe and InAs/GaSb quantum wells,\cite{Bernevig1757,PhysRevLett.107.136603,Konig2007a} but the small bulk gap arising from weak SOC makes the operating temperature very low and this limits its further applications.\cite{Li2015,C5CP00046G,C5RA10712A,Zhao2015a,Zhao2015b} To realize 2D topological insulators with a large band gap, most studies have focused on some heavy elements, such as Bi, which exhibit a strong SOC effect. The largest nontrivial bulk gap  (1.08eV) is found in Bi$_2$F$_2$ bilayer. The huge SOC gap in this material originates from the Bi $p_x$ and $p_y$ orbitals.\cite{Song2014,PhysRevB.90.085431} But also hexagonal Bi(111) bilayers have been realized and their time-reversal symmetry-protected edge states have been observed.\cite{PhysRevLett.109.016801} However, their topological nature is still debated. The search for other Bi-based QSH insulators is therefore interesting. 

%Currently, the nontrivial SOC gaps 0.33 eV or 0.29 eV are reported in 2D  Bi monolayer with 4-atom and 8-atom rings \cite{2053-1583-2-4-045010,Li2016} or 4-atom, 6-atom and 12-atom rings \cite{zhang2016bismuthylene}, and the band inversion happened between the $p_{xy}$ ($p_x$ and $p_y$) and $p_z$ orbitals.  However, one may ask if there are other dynamical stable structures of element Bi that simultaneously have a large nontrivial SOC gap?

In this work, we propose a new structure model with 4-atom, 6-atom, and 8-atom (4,6,8-atom) rings for the group V elements: Bi, Sb, and As. The formation energy of these proposed structures is lower than those of other reported 2D group V structures containing 4- and 8-atom rings.\cite{2053-1583-2-4-045010,Li2016} We find that their phonon spectra contain no imaginary frequency modes, indicating their dynamical stability. In the case of Bi, the calculated band structure suggests nontrivial topological properties with a relative large nontrivial bulk gap of 0.123 eV, resulting from a band inversion at the $\Gamma$ point.
% between the $p_{x,y}$ ($p_x$ and $p_y$) and $p_z$ orbitals due to the SOC effect.
The proposed Sb and As bilayers show large indirect band gaps with SOC, but these band gaps are trivial. 

We demonstrate that the properties of the proposed structures can be tuned by the number of 6-atom rings. For Bi, we show that the formation energy can be decreased while retaining the topologically nontrivial properties. The nontrivial SOC band gap can reach 0.373 eV, which is larger than that of other reported allotropes of Bi, except the hexagonal bilayer.\cite{2053-1583-2-4-045010,Li2016,zhang2016bismuthylene} %Moreover, the adjustability of our proposed new structures makes the structures more feasible in experiments.

%Especially, the new structure with Bi atoms is topological insulator with large nontrivial band gap 0.123 eV, and it is induced by SOC accompanied by the band inversion  between the $p_{xy}$ ($p_x$ and $p_y$) and $p_z$ orbitals. More importantly, we demonstrate that our new structures could be tuned by the 6-atom rings, and we also propose three tunable structures with Bi atoms retaining their topologically nontrivial properties. The nontrival SOC band gap can reach 0.373 eV which is larger than the ever reported Bismuth monolayers. Moreover, the adjustability of our proposed new structures makes the structures more feasible in experiments. 

\section{Computational Methods}
Our first-principles calculations are based on Density Functional Theory (DFT) with the projector augmented wave method as implemented in the Vienna \textit{ab initio} simulation package (VASP). \cite{PhysRevB.54.11169,PhysRevB.48.13115,Kresse1999} The generalized gradient approximation (GGA) in the form proposed by Perdew, Burke and Ernzerhof (PBE) \cite{PhysRevLett.77.3865} was chosen as the electron exchange-correlation functional. The structure relaxation including the atomic positions and lattice vectors was performed by the conjugate gradient (CG) scheme until the maximum force on each atom was less than 0.01 eV/\AA. The energy cutoff of the plane waves was set to 500 eV with an energy precision of $10^{-5}$ eV. The Brillouin zone (BZ) was sampled by using a 13 $\times$ 7$\times$ 1 $\Gamma$-centered Monkhorst-Pack grid. Phonon frequencies are calculated by the finite displacement method with the \textit{Phonopy} code. \cite{Togo2015a} 

The $Z_2$ topological invariants were obtained by calculating the Wannier Charge Centers (WCCs) and tracking the largest gap in the spectrum of the WCCs,\cite{PhysRevB.83.235401} which is equivalent to the computation of the Wilson loop.\cite{PhysRevB.84.075119} The explicit numerical computations were done with the Z2Pack code \cite{gresch2016z2pack} which combines the \textit{ab initio} calculations with the Wannier90 code.\cite{Mostofi2014} The surface state calculations are illustrated with an effective tight-binding Hamiltonian generated from the first-principles Wannier functions. The \textit{s} and \textit{p} orbitals of the Bi atoms from the first-principles wave functions are used as the initial trial orbitals. The iterative Green’s function method \cite{0305-4608-15-4-009} was used with the software package Wannier\_tools. \cite{wanntools}

\section{Results}

\subsection{Structure and Stability}

Due to the similarity of the proposed structures for the various investigated elements, we mainly focus on the structure model of Bi in this section. An example of such a structure is given in Fig.~1. Its lattice is rectangular, which is different from the hexagonal lattice of Bi(111) bilayers\cite{PhysRevLett.114.066402,Drozdov2014a,PhysRevB.90.165412,PhysRevLett.107.136805} and the square lattice of the recently proposed Bi bilayers consisting of 4,8-atom rings.\cite{2053-1583-2-4-045010,Li2016} The space group of the proposed orthorhombic crystals is \textit{Pccm} (or $D_{2h}^3$). There is a two-fold rotation, mirror, and inversion symmetry in this structure.  Along the $x$ direction, there are two kinds of arrangements of atomic rings. One is formed by the line along the center of 4(8)-atom rings while the other is along the center of the 6-atom rings, as indicated by the blue dashed line in Fig.~1(a). The two arrangements of atomic rings alternate along the $y$ direction and form a new type of Bi bilayer. Regarding the number of atomic rings, one 4-atom ring corresponds to one 8-atom ring and one 6-atom ring. Since the 4-atom rings always come in pairs with the 8-atom rings, our structure is denoted as a 4(8)-6 Bi bilayer in the following. As demonstrated below, such structures can be easily tuned by including more hexagons. A similar structure model can be applied to Sb and As.
%The top and side views are separately demonstrated in Fig.1(a) and (b). The 4, 8, 6-atom rings are labeled at the upper left corner of Fig.1(a). The unit cell is shown as  a rectangle in the center of Fig.1(a). There are twelve atoms with periodically buckled topology in the unit cell.  If we connect the geometrical centers of 4(8)-atom or  6-atom rings with blue dashed lines in the x direction as shown in Fig.1(a), we could see that in the y direction there is a periodically repetition of 4(8) and 6 connected lines. So our structure in this section is called as the 4(8)-6 structure. 
\begin{figure}[htb!]
\includegraphics[scale=0.58]{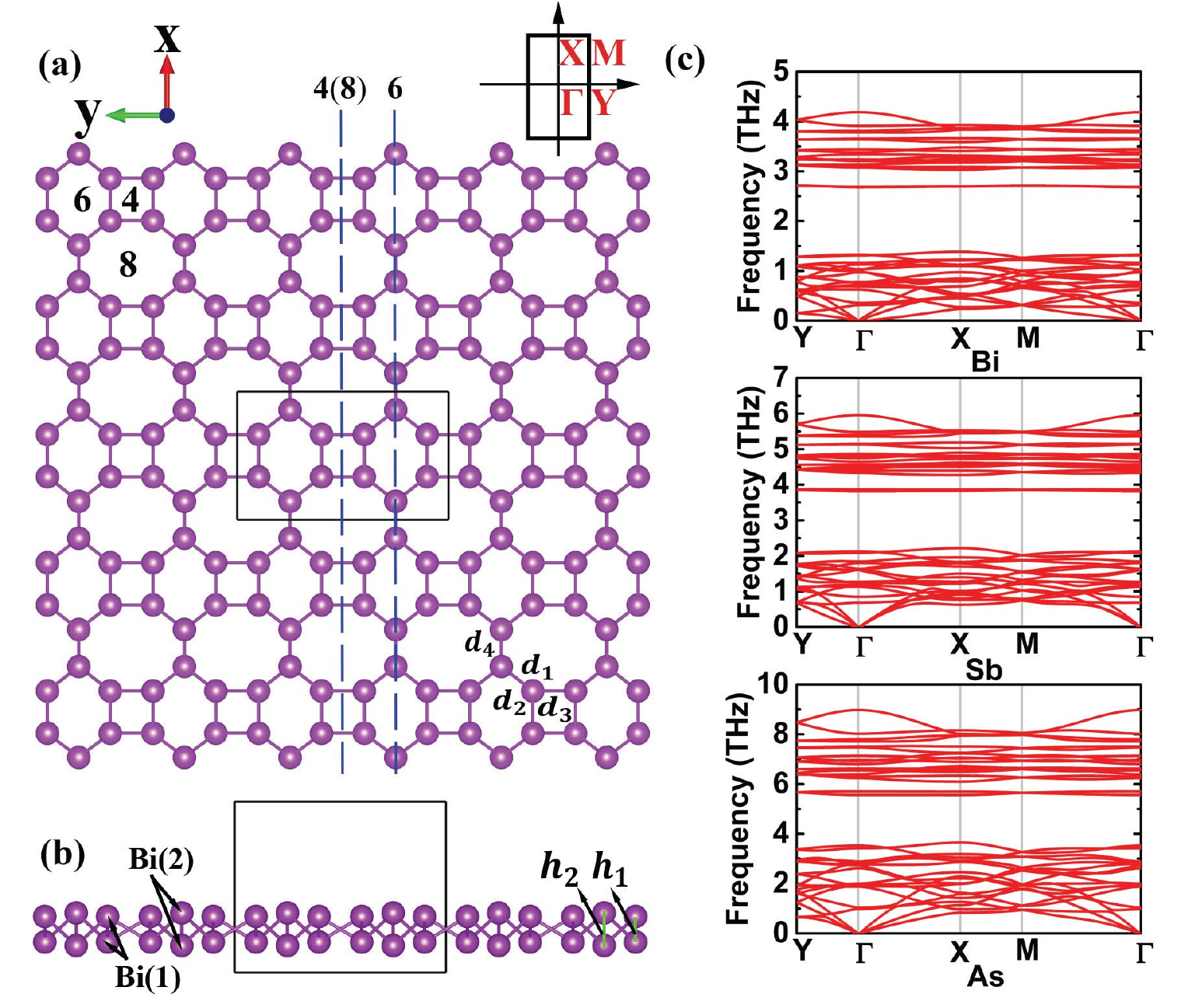}%
\caption{\label{486-structure} (a) Top view of the 4(8)-6 Bi bilayer: the rectangle indicates the unit cell, the blue dashed lines are the 4(8)-center-connected and 6-center-connected lines, and the inset in the upper right corner depicts the Brillouin zone (BZ) with the time-reversal invariant momenta (TRIM).  (b) Side view of the 4(8)-6 Bi bilayer. (c) Phonon spectra of 4(8)-6 Bi, Sb, and As bilayers.}
\end{figure}

 \begin{table*}[htb!]%[H] add [H] placement to break table across pages
 \caption{\label{structure_parameters} The optimized structure parameters of 4(8)-6 Bi, Sb and As bilayers. $a$ ($b$) is the lattice constant in the $x$ ($y$) direction; $h_1$ and $h_2$ is the buckling height as shown in Fig.~\ref{486-structure}(b); $d_{1,2,3,4}$ denotes the different bond lengths shown in Fig.~\ref{486-structure}(a).  $ \Delta$E is the formation energy defined by Eq. (1). }
 \begin{ruledtabular}
 \begin{tabular}{c c c c c c c c c c}
 elements & $a$ (\AA) & $b$ (\AA) & $h_1$ (\AA)  & $h_2$ (\AA) & $d_1$ (\AA) & $d_2$ (\AA) & $d_3$ (\AA) & $d_4$ (\AA) & $ \Delta$E (meV/atom) \\
 \hline
 Bi & 7.918 & 13.050 & 1.579 & 2.014 &  3.046 & 3.055 & 3.079 & 3.043 & 52.8 \\
 \hline
 Sb & 7.529 & 12.402 & 1.505 & 1.897 & 2.893 & 2.901 & 2.919 & 2.892 & 58.1\\
 \hline
 As  & 6.581 & 10.904 & 1.290 & 1.563 & 2.510 & 2.516 & 2.531 & 2.509 & 71.9 \\
 \end{tabular}
 \end{ruledtabular}
 \end{table*}

The optimized structure parameters for the 4(8)-6 bilayers of Bi, Sb, and As are listed in Table \ref{structure_parameters}. Due to the similarity of these  4(8)-6 bilayers, we focus on the Bi bilayer first. The lattice constant $a$ ($b$) in the $x$ ($y$) direction of Bi is 7.918 \AA\ (13.050 \AA). Although the lattice of the 4(8)-6 bilayer has different symmetry than the hexagonal and square Bi bilayers, the local arrangement of the neighboring atoms is similar.\cite{PhysRevLett.114.066402,Drozdov2014a,PhysRevB.90.165412,PhysRevLett.107.136805,2053-1583-2-4-045010,Li2016} One Bi atom forms a bond with three other Bi atoms that are all above or below than the position of the Bi atom in the $z$ direction. However, while there is only one kind of Bi atoms (one Wyckoff Position) in the hexagonal and square Bi bilayers, there are two kinds of Bi atoms (two Wyckoff Positions) in the 4(8)-6 structure, as illustrated in Fig.~1(b). Corresponding to these two kinds of Bi atoms, denoted as Bi(1) and Bi(2) in the following, there are two buckling heights, $h_1=1.579$ \AA\ and $h_2=2.014$ \AA. The buckling heights of the hexagonal (1.737 \AA) and square (1.757 \AA) Bi bilayer are in between the two heights of the 4(8)-6 Bi bilayer. 
%Although the buckled height is different from the hexagonal and square Bi bilayer, the bond length is similar with them. 
%the buckled height $h_1$ of Bi is 1.579\AA\ which is lower than the the buckled height in the honeycomb structure (1.737\AA) \cite{PhysRevLett.114.066402,Drozdov2014a,PhysRevB.90.165412,Liu2011} and 4(8) structure (1.757\AA),\cite{2053-1583-2-4-045010}  while the buckled height $h_2$ is 2.014\AA\ which is significantly larger than the the buckled height in the honeycomb structure and 4(8) structure. 
The Bi atoms in the 4(8)-6 bilayer are connected by four different bonds (see Fig.~1(a)) of which the lengths are shown in Table I. The length $d_1$ of the bond shared by the 6-atom and 8-atom rings is about 3.046 \AA\ which is practically the same as the bond length in the buckled hexagonal Bi bilayer (3.046 \AA).\cite{PhysRevLett.114.066402,Drozdov2014a,PhysRevB.90.165412,PhysRevLett.107.136805} The length $d_2$ of the bond shared by 4-atom and 6-atom rings is about 3.055 \AA\, which is slightly larger than $d_1$. The bond length $d_3$ shared by the 4-atom and 8-atom rings is about 3.079 \AA\ which is larger than the reported bond length (3.059 \AA) shared by the 4-atom and 8-atom rings in square Bi bilayer.\cite{2053-1583-2-4-045010} The bond length $d_4$ shared by the 8-atom and 8-atom rings is about 3.043 \AA\, which is nearly the same as the reported bond length (3.044 \AA) shared by the 8-atom and 8-atom rings in square Bi bilayer.\cite{2053-1583-2-4-045010} We can therefore conclude that the proposed structure is formed by an only slightly distorted combination of the square and hexagonal bilayer structures. Similar results are obtained for the 4(8)-6 Sb and As bilayers, although the structure parameters of Sb and As are smaller than that of 4(8)-6 Bi bilayer (see table I). This is in accordance with the general expectation that the lighter the atoms are, the smaller the structure parameters become.

%We could see that the heavier the elements, the structure parameters would be more larger.  The lattice constant a (b) in the x (y) direction of Bi is 0.38\AA\ (0.55\AA) larger than that of Sb, while the lattice constant in the x (y) direction of Sb is  0.95\AA\ (1.50\AA) larger than that of As. The condition with the averaged buckled height is similiar. The averaged buckled height (h) of Bi is  0.09\AA\  larger than that of Sb, and  the averaged buckled height of Sb is 0.25\AA\  larger than that of As. Here, the averaged buckled heights in 4(8)-6 structures with Bi, Sb, As is smaller than the repored buckled heights in 4(8) structures.\cite{Li2016}

Next, let us focus on the stability of the 4(8)-6 bilayers. To this end, we define the formation energy with respect to the hexagonal bilayer as follows:
\begin{equation}
\Delta E = (E_{total}-N_{atom} \times \mu_{atom})/N_{atom},
\end{equation}
where $E_{total}$ is the total energy of the 4(8)-6 bilayer, $N_{atom}$ is the total number of atoms in the crystal structure and $\mu_{atom}$ is the energy per atom calculated for the hexagonal honeycomb structure. 
%The energy of Bi atom (52.8 meV/atom) is the smallest, and it is 5.3 meV/atom smaller than that of Sb atom while the Sb atom is 13.8 meV/atom smaller than that of As atom. The lighter the atom is, the formation energy is larger which indicates the structure becomes more unstable. 
Starting from Bi, $\Delta E$ increases with decreasing atomic number.
The formation energy for Bi (52.8 meV/atom) is the smallest, while $\Delta E$ for Sb and especially As becomes somewhat larger. For comparison, the formation energy of a 4(8)-6 P bilayer was also calculated, and it was found to be even larger (76.8 meV/atom). Furthermore, the phonon spectrum of the P bilayer exhibits imaginary frequency modes, indicating its dynamical instability. Therefore, we will not consider the 4(8)-6 P bilayer in this work. 
For Bi, the formation energy of the 4(8)-6 Bi bilayer (52.8 meV/atom) is significantly smaller than that of a square Bi bilayer (80.6 meV/atom).\cite{2053-1583-2-4-045010} A similar behavior can be observed in 2D C allotropes, where the formation energy decreases with increasing number of C hexagons. In our case, the 4(8)-6 bilayers should be more stable than the reported square bilayers due to the larger number of hexagons.\cite{2053-1583-2-4-045010,Li2016}
To investigate the dynamical stability of the 4(8)-6 bilayers, their phonon spectrum along the high symmetry lines in the BZ is calculated from first principles using a supercell approximation (see Fig.~\ref{486-structure}(c)). It can be seen that the 4(8)-6 bilayers of Bi, Sb, and As are all dynamically stable, because no imaginary frequency modes are observed in their phonon spectrum.

\subsection{Electronic Band Structure}

\begin{figure}[htb!]
\includegraphics[scale=0.57]{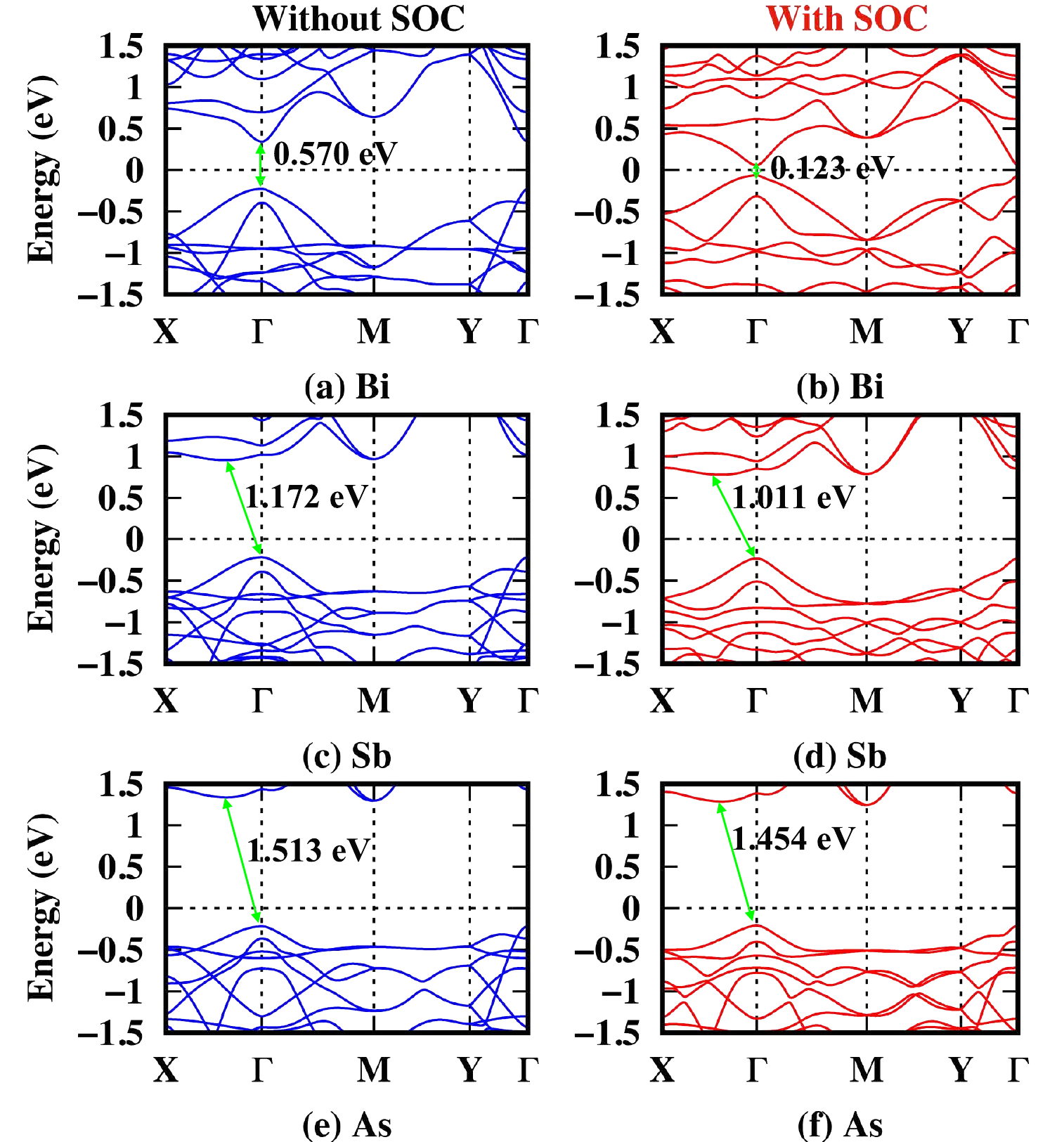}%
\caption{\label{486-BS} (a), (c) and (e) are the band structures of  4(8)-6 Bi, Sb, and As bilayers without SOC; (b), (d) and (f) are the band structures with SOC. The band gaps are given in the figure.}
\end{figure}

As mentioned in the introduction, 2D materials often have interesting electronic properties. The electronic band structure of the investigated  4(8)-6 bilayers is shown in Fig.~\ref{486-BS}. The left figures ((a), (c), and (e)) were calculated without SOC and the right ones ((b), (d), and (f)) were calculated with SOC. Let us consider the 4(8)-6 Bi bilayer first. It has a direct band gap of 0.570 eV at the $\Gamma$ point without SOC, which is similar to the hexagonal and square Bi bilayer.\cite{PhysRevB.90.165412,2053-1583-2-4-045010,Li2016}  With inclusion of SOC, the 4(8)-6 Bi bilayer retains its direct band gap, contrary to the case of hexagonal and square Bi bilayers which get an indirect band gap with SOC. The value of the direct band gap becomes 0.123 eV, which is smaller than that of the hexagonal and square Bi bilayer.

On the other hand, the Sb/As 4(8)-6 bilayers have indirect band gaps with and without SOC. This is similar to the hexagonal Sb/As bilayers, but different from the square ones which have a direct band gap  both with and without SOC.
The 4(8)-6 Sb bilayer has a large indirect SOC band gap of 1.011 eV, which is slightly smaller than the direct band gap (1.13eV) of the square Sb bilayer.\cite{Li2016} 
The 4(8)-6 As bilayer has the largest indirect SOC band gap (1.454 eV) of the three structures, but it is still smaller than the direct gap of the square As bilayer (1.71 eV).\cite{Li2016} Although, 4(8)-6 Bi, Sb, and As bilayers have substantially different band gaps, the regular pattern that the gap values increase with decreasing atomic number is similar to the hexagonal/square group V bilayers.
%For example in Bi 4(8)-6 structure \cite{2053-1583-2-4-045010} or Bi honeycomb structure \cite{PhysRevB.90.165412}, the direct gap appeared in the condition of no SOC, while the indirect gap appeared in the condition of  SOC. On the other hand, the Sb and As 4(8)-6 structures behave direct band gaps in the condition of no SOC and  indirect band gaps in the condition of SOC. This is quite similar with the Sb and As 4(8) structures.\cite{Li2016} However, when it comes to Sb and As buckled honeycomb structures, the large indirect band gaps were reported under both of the condition without SOC and with SOC.\cite{Zhao2015,Wang2015,zhang2015atomically} The As 4(8)-6 structure has the largest gaps of the three structures, i.e., 1.513 eV without SOC and 1.454 eV with SOC. However, the results are quite smaller than that of the As 4(8) structure \cite{Li2016} and also the buckled honeycomb structure.\cite{zhang2015atomically} The Sb 4(8)-6 structure has smaller band gaps than that of As: 1.172 eV without SOC and 1.011eV with SOC. The values of the band gaps are smaller than that of Sb 4(8) structure \cite{Li2016} and that of Sb buckled honeycomb structure. \cite{Zhao2015,Wang2015} 
%As the reported band gaps in the 4(8) structures would become smaller when the atom number becomes larger in Group-V elements,\cite{Li2016} the situation is the same with our case. 

\subsection{Topological Properties}

Since the electronic band gaps with SOC of hexagonal and square Bi bilayers have been shown to be nontrivial,\cite{PhysRevB.90.165412,2053-1583-2-4-045010,Li2016} we investigate the topological properties of the 4(8)-6 Bi bilayers here. Note that structures containing Bi atoms are often reported to be topologically nontrivial due to the strong intrinsic SOC of Bi. 
%In the following, we will show that the Bi 4(8)-6 in our work is also topological insulator with nontrivial SOC band gap. The nontrivial SOC band gap in Bi 4(8)-6 structure is smaller than that of Bi 4(8) structures,\cite{2053-1583-2-4-045010,Li2016} but the nontrivial SOC band gap could be tuned to 0.373 eV by the number of 6-atom rings which is larger than that of Bi 4(8) structures\cite{2053-1583-2-4-045010,Li2016} as will be discussed in Section V.
%The Bi 4(8)-6 structure has the smallest band gaps of the three structures at the $\Gamma$ point: 0.570 eV without SOC and 0.123 eV with SOC. On the other hand, the reported 4(8) structures with Bi, Sb, As have the same rules: As 4(8) structure has the largest gaps and the gaps will decrease as it comes to Sb and Bi.\cite{Li2016}
To investigate the topological properties of the 4(8)-6 Bi bilayer, we calculated  the $Z_2$ topological invariant by tracking the largest gap in the spectrum of the WCCs. In addition to the Wilson like methods, we also calculated the $Z_2$ invariants by parity analysis because the 4(8)-6 Bi bilayer has inversion symmetry. Following Fu \textit{et al.},\cite{PhysRevB.76.045302} the $Z_2$ topological invariant ($\upsilon$) in systems with time-reversal symmetry and inversion symmetry can be obtained by:
\begin{equation}
(-1)^{\upsilon}=\prod^4_{i=1}\delta(K_i), \delta(K_i)=\prod^N_{m=1}\xi^i_{2m},
\end{equation}
with $K_i$ the TRIMs, $\xi=\pm 1$ the parity eigenvalue of the wave function, $\delta(K_i)$ the product of the parity eigenvalues at the TRIM, and $N$ the total number of degenerate occupied bands. In our case $K_i$ is $\Gamma$, X, Y, or M. The $Z_2$ topological invariant of the 4(8)-6 bilayer equals 1, which proves its nontrivial nature. We list the results of the parity eigenvalues in Table II. It is seen that only the Bi bilayers have a nontrivial topological invariant ($\upsilon=1$), with the only difference in parity eigenvalues between 4(8)-6 Bi and Sb or As bilayers at the $\Gamma$ point.  The parity eigenvalue of -1 at the $\Gamma$ point of the 4(8)-6 Bi bilayer suggests that there is a band inversion at this TRIM (see below). Similar to the hexagonal and square bilayers, only the Bi bilayer has a topologically nontrivial band gap, while Sb and As bilayers have trivial gaps.\cite{PhysRevB.90.165412,Li2016}

\begin{table}[htb!]%[H] add [H] placement to break table across pages
 \caption{\label{parity eigenvalues} The parity eigenvalues at the four TRIMs ($\Gamma$, X, Y, M) and the $Z_2$ topological invariants ($\upsilon$) of the 4(8)-6 Bi, Sb, and As bilayers.}
 \begin{ruledtabular}
 \begin{tabular}{c c c c c c}
 element & $\Gamma$ & X & Y & M & $\upsilon$ \\
\hline
Bi & -1 & 1 & -1 & -1 & 1 \\
\hline
Sb & 1 & 1 & -1 & -1 & 0\\
\hline
As & 1 & 1 & -1 &  -1 & 0 \\
 \end{tabular}
 \end{ruledtabular}
\end{table}

\begin{figure}[htb!]
\includegraphics[scale=0.345]{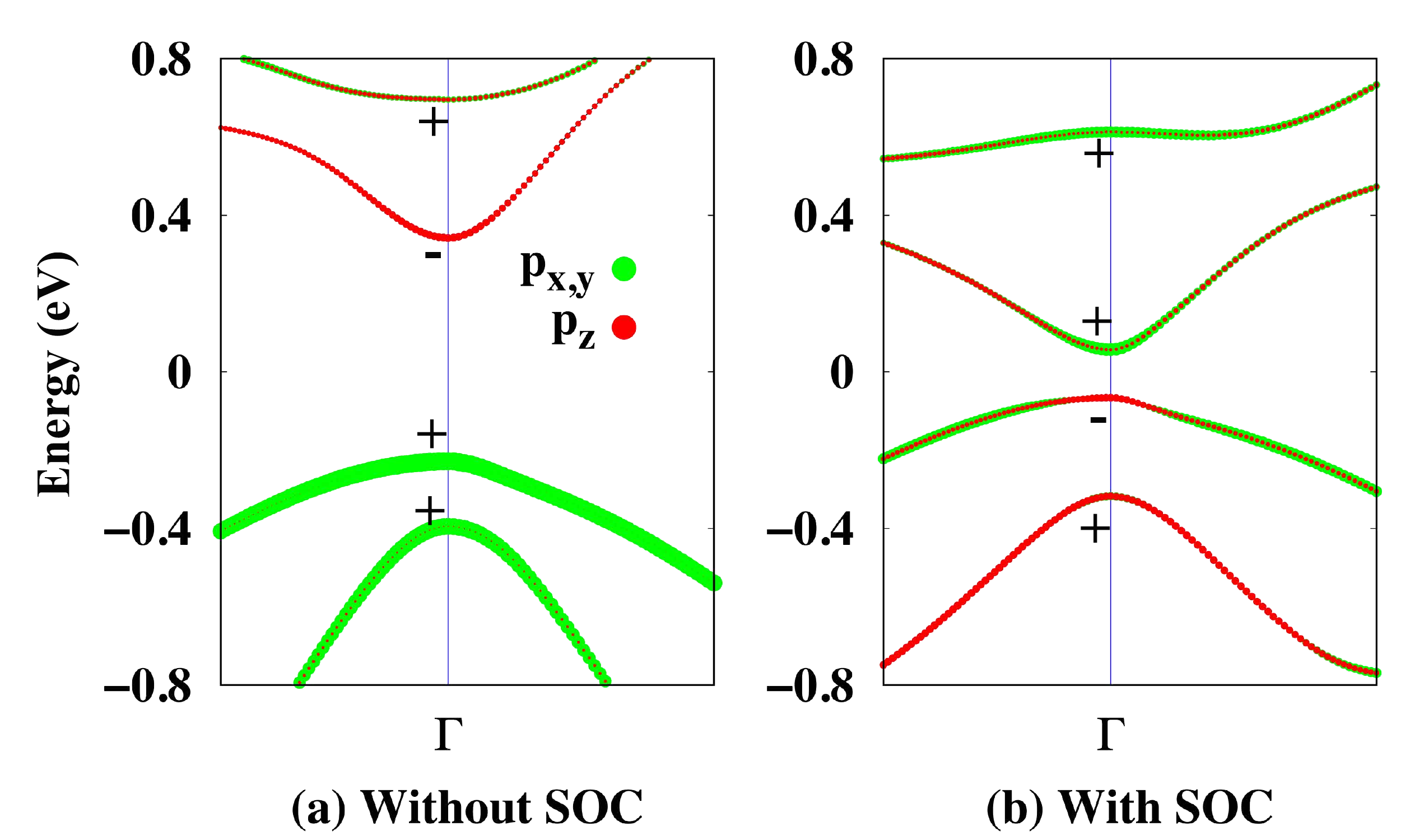}%
\caption{\label{486-structure} The orbital-projected band structures of the 4(8)-6 Bi bilayer: (a) without SOC and (b) with SOC. The symbol size indicates the contribution weight: larger dot means higher contribution while smaller one indicates lower contribution. $\pm$ indicates an even or odd parity eigenvalue. Green: $p_{x,y}$ ($p_x$ and $p_y$) orbitals; red: $p_z$ orbitals.}
\end{figure}
\begin{figure}
\includegraphics[scale=0.075]{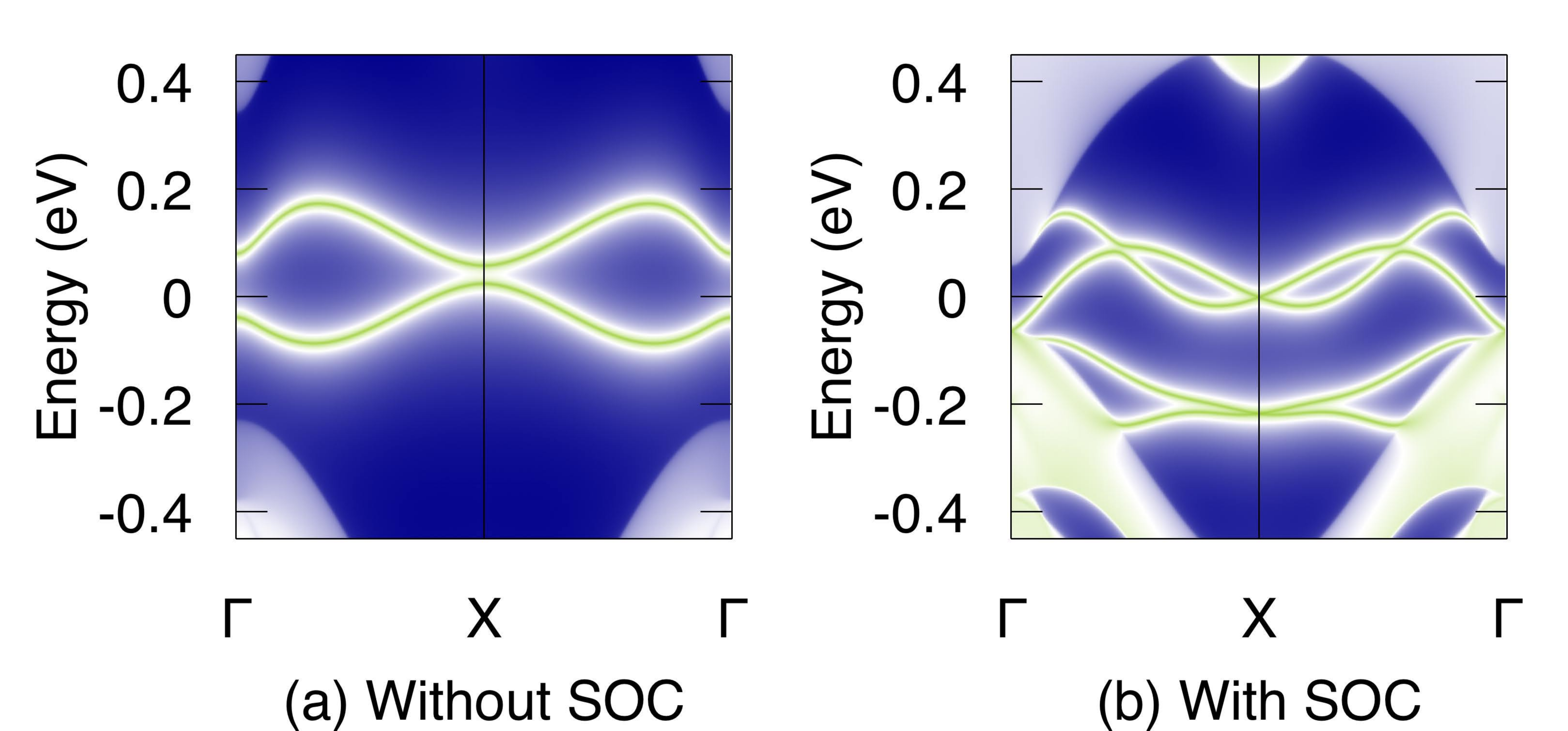}%
\caption{\label{486-structure} The edge states of 4(8)-6 Bi bilayer: (a) without SOC; and (b) with SOC.}
\end{figure}
\begin{figure*}[htb!]
\includegraphics[scale=0.8]{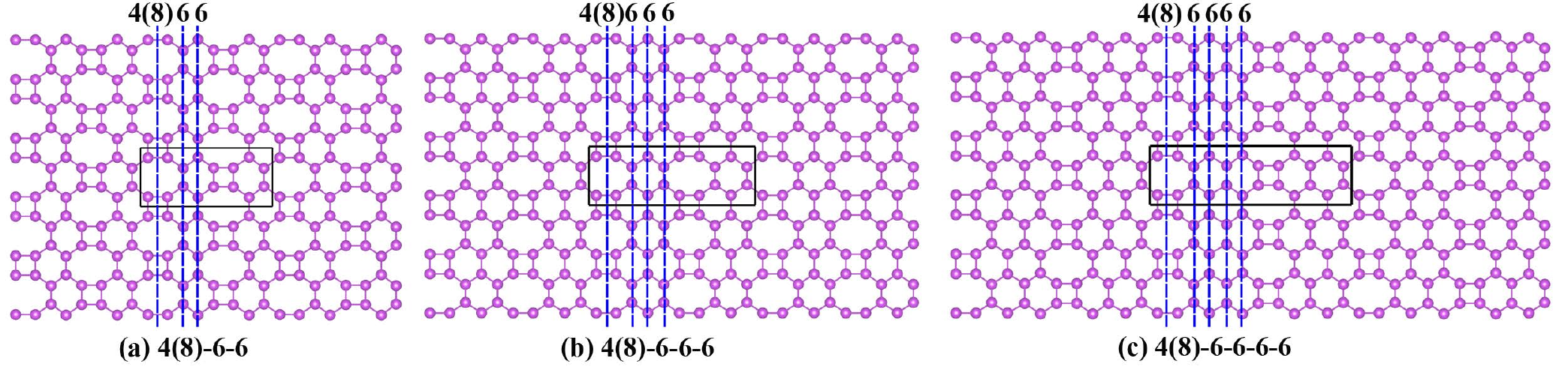}%
\caption{\label{tunable_structures} The tunable structures: (a) 4(8)-6-6; (b) 4(8)-6-6-6; (c) 4(8)-6-6-6-6 Bi bilayers. The blue dashed lines are the 4(8)-center-connected and 6-center-connected lines.}
\end{figure*}
To find the origin of the topologically nontrivial nature of the 4(8)-6 Bi bilayer, we investigate the band inversion by the orbital-projected band structures as shown in Fig.~3. We can see that SOC plays an important role in the inversion of the states with $p_{x,y}$ and $p_z$-orbital character: without SOC at the $\Gamma$ point, the $p_{x,y}$ orbitals contribute the most to the highest occupied band and the $p_z$ orbitals contribute the most to the lowest unoccupied band, but the situation is reversed when including SOC. This is also confirmed by the reversal of parity eigenvalues between the highest occupied band and the lowest unoccupied band at the $\Gamma$ point. Note that the observed band inversion found here is similar to the case of hexagonal and square Bi bilayers whose band inversion also occurs between the $p_{x,y}$ and $p_z$ orbitals.\cite{PhysRevB.90.165412,2053-1583-2-4-045010,Li2016} The band inversion indicates that there must be a gap closing with  corresponding formation of a Dirac cone when continuously turning on the SOC.\cite{2053-1583-2-4-045010,li2016gallium}

Besides the nonzero $Z_2$ topological invariant and the observed band inversion, the existence of gapless edge states is another prominent feature of QSH insulators.
According to the bulk-edge correspondence in topological insulators, a nontrivial topological invariant ($\upsilon=1$) indicates the presence of topologically protected edge states at the edges of the material. The calculated edge states of the 4(8)-6 Bi bilayer are shown in Fig.~4. In Fig.~4(a), it is seen that there are edge states in the band gap without SOC, but these edge states do not bridge the band gap which indicates that they are trivial. However, including SOC, there are two oppositely propagating gapless edge states appearing in the bulk gap that connect the conduction and valence bands and which cross at the TRIM (X point) as shown in Fig.~4(b).

\subsection{Structural Tunability}

As we discussed above, the formation energy decreases with increasing number of 6-atom rings in the system. In this section we investigate how the properties of the 4(8)-6 Bi bilayer can be tuned by changing the number of hexagons. This is done in a systematic way  by increasing the number of 6-connected-lines which connect the centers of 6-atom rings in the $x$ direction. Taking Bi as an example, we show three such structures in Fig.~5. As indicated by the blue dashed lines, the number of 6-connected-lines is 2, 3, and 4 in these structures and they are correspondingly named 4(8)-6-6, 4(8)-6-6-6 and 4(8)-6-6-6-6. The calculated structure parameters of the three structures are shown in Table III.  Due to the similar structure along the $x$ direction compared to 4(8)-6, the lattice constants of the expanded structures in the $x$ direction are similar (7.919 \AA, 7.778 \AA, 7.708 \AA, and 7.661 \AA) and are in between those of the square and hexagonal structure.  We can easily understand this: the larger the number of the 6-atom rings becomes, the closer to the hexagonal Bi bilayer the expanded structure gets. There is now a larger variety of different Bi atoms, corresponding to various buckling heights, so we compare the averaged buckling heights. These are 1.724 \AA, 1.735 \AA, 1.734 \AA, and 1.734 \AA, respectively.  Note that these averaged buckling heights are larger than the buckling height of 1.71\AA\ as found in the hexagonal Bi bilayer\cite{PhysRevLett.107.136805,PhysRevB.87.235420} and smaller than 1.76\AA\ in the square Bi bilayer.\cite{2053-1583-2-4-045010,Li2016}
\begin{figure}[htb!]
\includegraphics[scale=0.3]{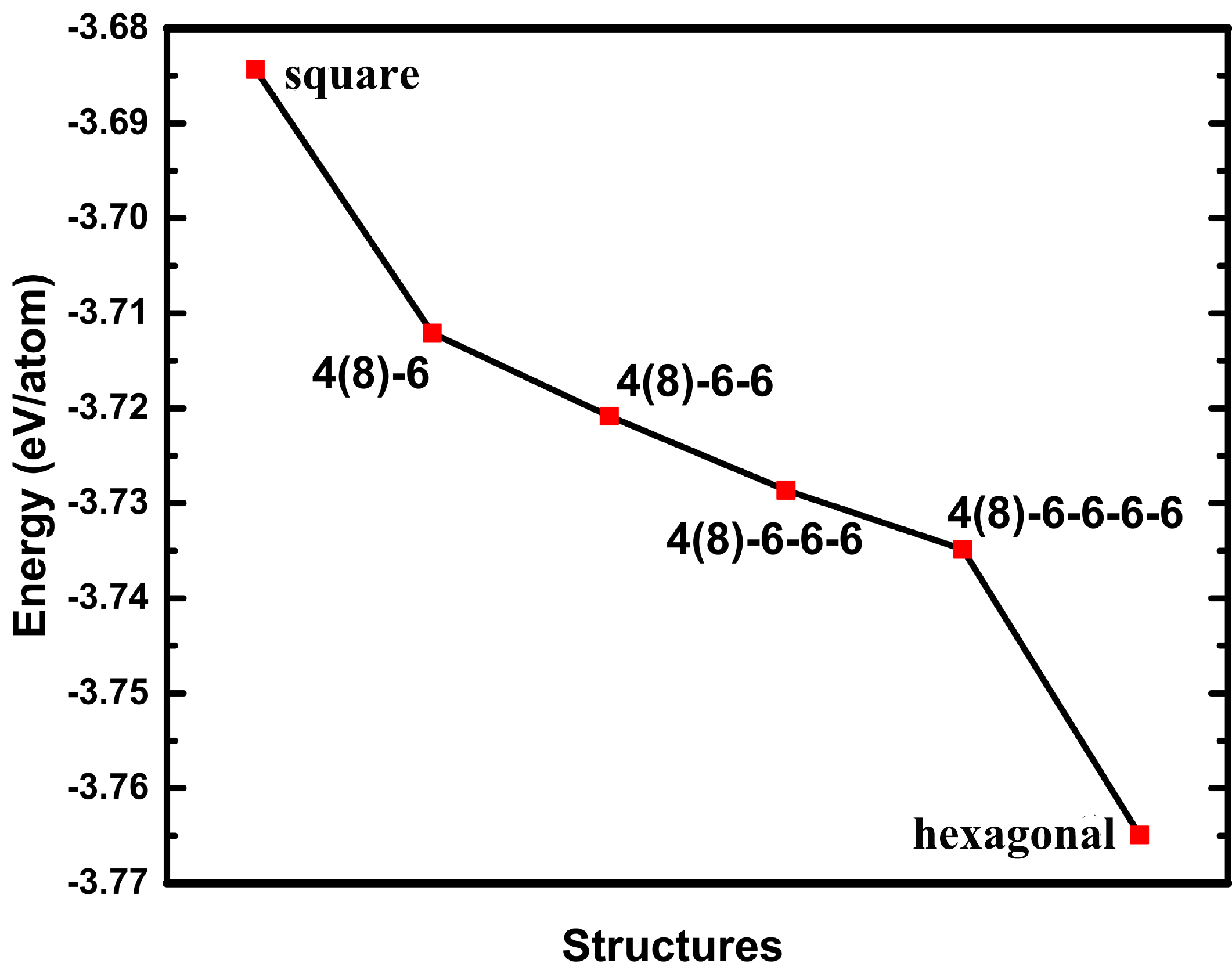}%
\caption{\label{tunable structures} The energy per atom of 4(8)-6, 4(8)-6-6, 4(8)-6-6-6, 4(8)-6-6-6-6 Bi bilayers.}
\end{figure}

\begin{table}[htb!]%[H] add [H] placement to break table across pages
 \caption{\label{structure parameters} The optimized structure parameters of 4(8)-6-6, 4(8)-6-6-6, and 4(8)-6-6-6-6 Bi bilayers. $a$ ($b$) is the lattice constant in the $x$ ($y$) direction; $h$ is the averaged buckling height.}
 \begin{ruledtabular}
 \begin{tabular}{c c c c}
 structures & $a$ (\AA) & $b$ (\AA) & $h$ (\AA) \\
 \hline
 4(8)-6-6 & 7.778 & 17.493 & 1.735 \\
 \hline
 4(8)-6-6-6 & 7.708 & 21.834 & 1.734 \\
 \hline
 4(8)-6-6-6-6  & 7.661 & 26.188 & 1.734\\
 \end{tabular}
 \end{ruledtabular}
\end{table}

\begin{figure}[htb!]
\includegraphics[scale=0.54]{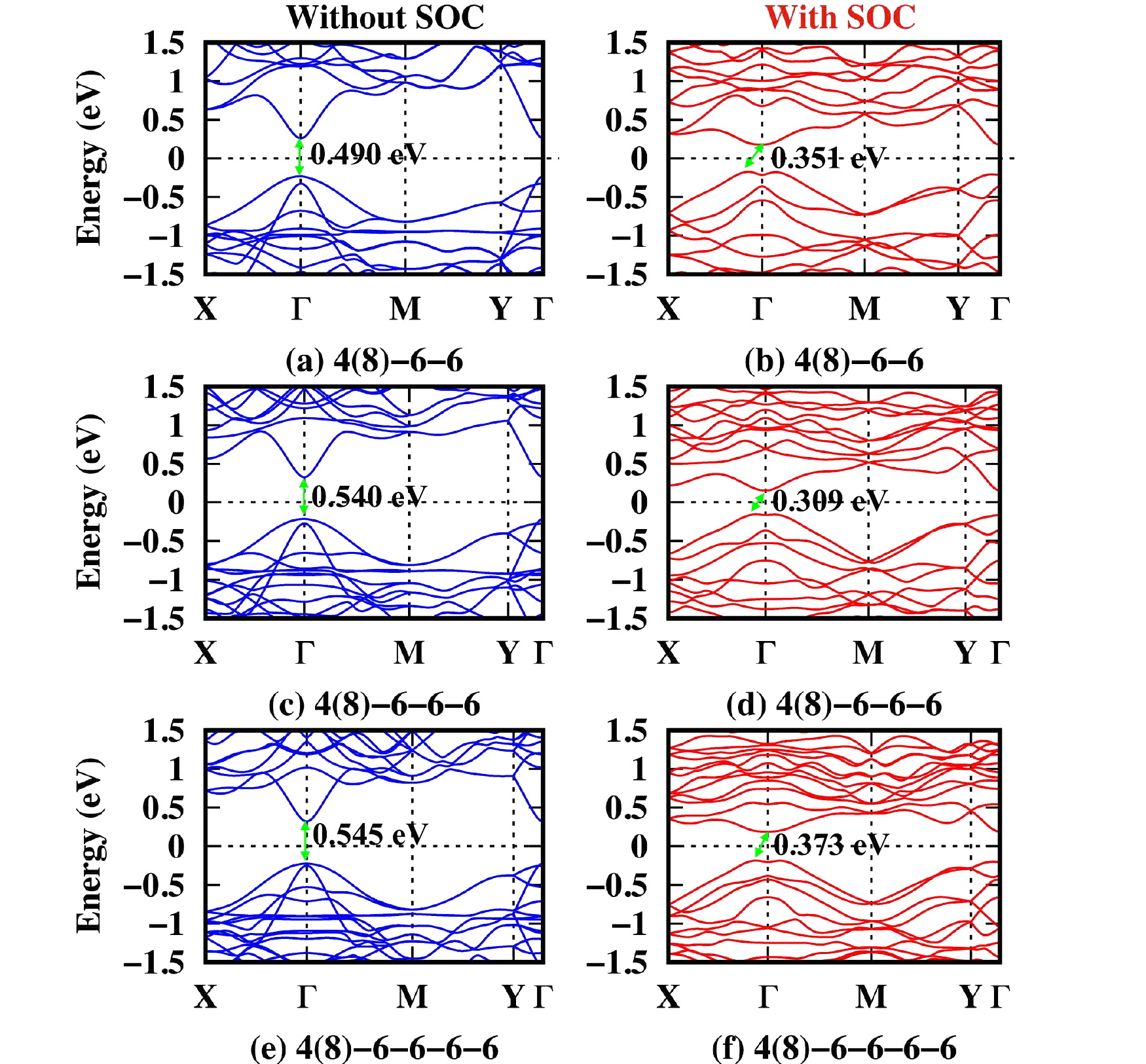}%
\caption{\label{tunable structures} (a), (c), and (e) are the band structures of the 4(8)-6-6, 4(8)-6-6-6, and 4(8)-6-6-6-6 Bi bilayers without SOC; (b), (d), and (f) are the band structures with SOC. The band gaps are marked in the figure.}

\end{figure}
\begin{figure}[htb!]
\includegraphics[scale=0.035]{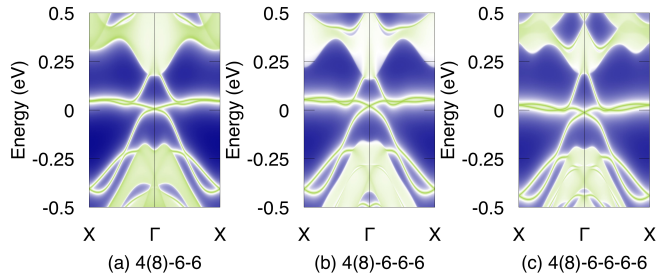}%
\caption{\label{tunable structures} The nontrival edge states of the structures: (a) 4(8)-6-6, (b)4(8)-6-6-6, and (c) 4(8)-6-6-6-6 Bi bilayers.}
\end{figure}

We also calculated the energy per atom in all these structures and compare them to the hexagonal and square Bi bilayers which have been studied before.\cite{PhysRevLett.107.136805,PhysRevB.87.235420,2053-1583-2-4-045010,Li2016}  As shown in Fig.~6,  the hexagonal Bi bilayer has the lowest energy which implies it is the most stable of all these structures. On the other hand, the square Bi bilayer has the largest energy of all. The energy per atom of the new proposed structures varies monotonously between these two limiting structures and converges to the energy of the hexagonal bilayer as more hexagons are included. 
 
The electronic band structures of the 4(8)-6-6, 4(8)-6-6-6, 4(8)-6-6-6-6 Bi bilayers are shown in Fig.~7. As shown in Fig. 7 (a), (c) and (e), the direct band gaps calculated without SOC of the three structures are 0.490 eV, 0.540 eV and 0.545 eV, which is smaller than that of 4(8)-6 Bi bilayer. After inclusion of SOC, as shown in Fig. 7 (b), (d) and (f),  the band gaps of the three structures are  0.351 eV, 0.309 eV and 0.373 eV. Note that the band gaps become indirect in contrast to the direct band gap of the 4(8)-6 Bi bilayer. Compared with the hexagonal/square Bi bilayer,\cite{PhysRevB.90.165412,2053-1583-2-4-045010,Li2016}, only the 4(8)-6 Bi bilayer with SOC has a direct gap. Calculations of the $Z_2$ topological invariant show that all the new structures are topological insulators induced by SOC. Moreover, the indirect nontrivial SOC gap becomes larger upon inclusion of more Bi hexagons and approaches the gap of the hexagonal Bi bilayer.\cite{2053-1583-2-4-045010,Li2016, zhang2016bismuthylene}The nontrivial edge states of 4(8)-6-6, 4(8)-6-6-6, 4(8)-6-6-6-6 Bi bilayers are shown in Fig.~8. The topological edge states of the three structures appear to be very similar.

In general, we can conclude that as the number of the 6-atom rings increases, the energy, electronic structure, and nontrivial band gap approach those of the hexagonal Bi bilayer. At the same time, the lines of 4-atom and 8-atom rings can be regarded as line defects in the hexagonal Bi bilayer. By introducing such line defects, it is possible to tune the properties of the hexagonal Bi bilayer.

\section{Conclusions}
Using first-principles calculations, we propose a new stable 4(8)-6 model for the group V elements (Bi, Sb, As), which enrich the family of 2D materials. Their formation energy compares favorably to square bilayers while our phonon calculations confirm their dynamical stability. The trivial (Sb, As) and nontrivial (Bi) band gap make the group V 4(8)-6 structures promising candidates for applications in future nanodevices. The nontrivial topological phase of the Bi 4(8)-6 structure was demonstrated by the calculations of $Z_2$ topological invariant, band inversion and the edge states. Interestingly, the Bi 4(8)-6 structure has a direct band gap in contrast to the previously studied square and hexagonal Bi bilayers. Moreover, the 4(8)-6  model allows for property tuning by changing the ratio of hexagons in the structure. In the case of Bi, we investigated 3 such models, namely 4(8)-6-6, 4(8)-6-6-6, and 4(8)-6-6-6-6, and found that they are all topological insulators.  As the number of the hexagons increases, the energy, electronic structure, and nontrivial band gap approach those of the hexagonal Bi bilayer. In the dilute limit, the lines of 4-atom and 8-atom rings can be regarded as line defects in the hexagonal Bi bilayer that can be used to tune the properties of the hexagonal Bi bilayer.

% The flexible tuning in our proposed new 2D materials make them more feasible in experiments. 
%Our proposed dynamical stable group-V 4(8)-6 structures enrich the new families of 2D materials. The nontrivial topological properties of Bi 4(8)-6 structures make the group-V 4(8)-6 structures more promising potential applications in future nanodevices. The physics of the nontrivial topological phase of Bi 4(8)-6 structure is illustrated by the calculations of  $Z_2$ topological invariant, parity eigenvalues, orbital-projected band structures and edge states. Moreover, the tunable structures based on Bi 4(8)-6 structure are demonstrated as Bi 4(8)-6-6, 4(8)-6-6-6, 4(8)-6-6-6-6, and they  are all topological insulators. The large nontrivial SOC band gap 0.373 eV in Bi 4(8)-6-6-6-6 structure is larger than any other reported allotropes of Bi monolayer. The flexible tuning in our proposed new 2D materials make them more feasible in experiments.

\begin{acknowledgments}
%This work including a part of the computational resources and services was supported by MOST (Grant No. 2016YFA0301604), NSFC (No. 11574008), and Thousand-Young-Talent Program of China. This work was also supported by the Fonds Wetenschappelijk Onderzoek (FWO-Vl). Another part of the computational resources and services used in this work were provided by the VSC (Flemish Supercomputer Center), funded by the Research Foundation - Flanders (FWO) and the Flemish Government – department EWI.
This work is supported by the MOST (Grant No. 2016YFA0301604), NSFC (No. 11574008), Thousand-Young-Talent Program of China, and Fonds Wetenschappelijk Onderzoek (FWO-Vl). The computational resources and services used in this work were provided by the VSC (Flemish Supercomputer Center), funded by the Research Foundation - Flanders (FWO) and the Flemish Government – department EWI.

\end{acknowledgments}

\bibliography{aps_mau.bib}

\end{document}